\documentclass[aps,preprint,prd,nofootinbib]{revtex4}

\usepackage{graphicx}
\usepackage{psfrag}

\usepackage{subfigure}
\usepackage{array}
\usepackage{graphicx}
\usepackage{amsmath}
\usepackage{amssymb}
\usepackage{mathrsfs}
\usepackage{bm}
\usepackage{slashed}
\usepackage{threeparttable}
\usepackage{multirow}

\usepackage[colorlinks=true, linkcolor=blue, citecolor=blue, urlcolor=blue]{hyperref}

\usepackage[amsmath,thmmarks,hyperref]{ntheorem}
\usepackage{soul}

\allowdisplaybreaks[4]

\begin{document}

\title{Mass spectra of heavy pseudoscalars using instantaneous Bethe-Salpeter equation with different kernels}

\author{Wei Li$^{1,2}$}
\author{Ying-Long Wang$^{1,2}$}
\author{Tai-Fu Feng$^{1,2}$}
\author{Guo-Li Wang$^{1,2}$ \footnote{Corresponding author}}

\address{1, Department of Physics, Hebei University, Baoding 071002, China\\
2, Key Laboratory of High-precision Computation and Application \\  of Quantum Field Theory of Hebei Province, Baoding, China}

\baselineskip=20pt

\begin{abstract}
We solved the instantaneous Bethe-Salpeter equation for heavy pseudoscalars in different kernels, where the kernels are obtained using linear scalar potential plus one gluon exchange vector potentials in Feynman gauge, Landau gauge, Coulomb gauge and time-component Coulomb gauge. We obtained the mass spectra of heavy pseudoscalars, and compared the results between different kernels, found that using the same parameters we obtain the smallest mass splitting in time-component Coulomb gauge, the similar largest mass splitting in Feynman and Coulomb gauges, middle size splitting in Landau gauge.
\end{abstract}

\maketitle

\section{Introduction}
After the year 2003, mass spectra of hadrons have been greatly enlarged, many new particles were discovered in experiments by LHCb, BaBar, Belle, BESIII, ATLAS, CMS and D0 collaborations. For example, $X(3872)$ \cite{3872}, $D(2550)$ \cite{2550}, $D_J(2580)$ \cite{2580}, $B_c(2S)$ \cite{bc2s}, $D^*_{s1}(2700)$, $D^*_{sJ}(2860)$ \cite{2700}, $\eta_b(2S)$ \cite{mizuk}, $B_J(5840)$ \cite{5840}, $Z_c(3900)^+$ \cite{3900}, etc. As summary of these new particles, see the review papers \cite{zhu,zhu2,olsen,swanson}.

Theoretically, the Quantum Chromodynamics (QCD) is the underlying theory of strong interaction responsible for the binding of quarks in hadron, but because of the non-perturbative nature of QCD at low energy, it can not be used to study the hadrons directly. Thus,  effective models based on QCD like the constitute quark model, lattice QCD, QCD sum rules, chiral quark model, etc, are introduced to study hadron properties.

Quark model is one of the most successful theory in studying the hadron spectroscopy. Within the model, various possible potentials are chosen based on the aspects of pragmatism to study the mass spectra of hadrons, for example, the logarithmic potential \cite{quigg,eichten}, Coulomb-plus-linear potential (Cornell potential) \cite{eichten1,godfrey}, power-law potential \cite{martin,eichten}, Buchmuller-Tye potential \cite{buchmuller}, confining vector potentials \cite{olsson}, one gluon exchange with mixture of long-range vector and scalar linear confining potentials \cite{ebert},
Cornell potential with relativistic corrections \cite{eichten2,godfrey1,barnes}, Lorentz gauge kernels \cite{lucha}, etc. We have known lots of information about the hadron spectra \cite{godfrey}, and believed that the potential model is very successful until the year 2003. Since then, we found that many new particles can not be accommodated by the traditional potential model, which means we still need paying more attention on hadron spectra as well as the potentials.

We should note that to a certain extent, most of the potential models are based on Bethe-Salpeter (BS) equation \cite{BS}, or its different approximations including Schrodinger equation. So in this paper, based on solving the instantaneous BS equation, we will study the mass spectra of heavy pseudoscalar, and focus on various potentials caused by different gauges. BS equation is a relativistic method describing bound state, it is also one of the relativistic quark model. Though it is based on quantum field theory, it can not give the information of interaction kernel between quarks, so we have to determine it considering the theory of QCD. It has several scales in QCD, in small range, there is the asymptotic freedom effect between quarks, then the perturbative QCD is welcome, while in large range, there is the color confinement, and the linear potential is found. In perturbative QCD, the strong interaction is transferred by exchanging gluon, while we know that the propagator of gluon can choose different gauges, so this paper, we will study the mason mass spectra in different kernels caused by different gauges, and see the differences.

In section II, we give a brief review of the Bethe-Salpeter equation and Salpeter equation, the latter is the instantaneous version of the former. In Sec. III, we show how we obtain the vector potentials using the single-gluon exchange in different gauges. In Sec. IV, the relativistic wave function is given. The results of mass spectra of pseudoscalar using different kernels and a short discussion are present in Sec. V.

\section{Bethe-Salpeter Equation and Salpeter Equation}

\begin{figure}[htb]
\begin{picture}(470,600)(125,490)
\put(0,0){\includegraphics{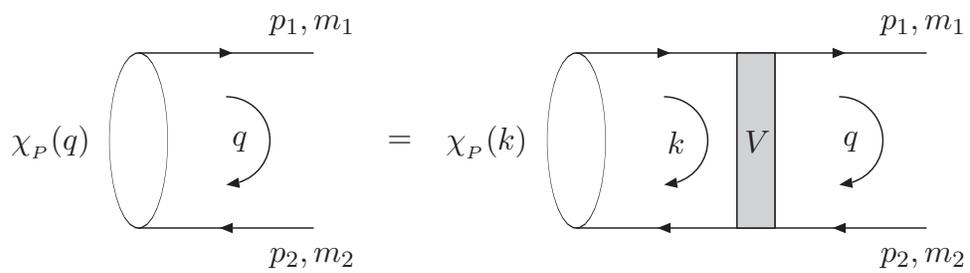}}
\end{picture}
\caption{Bethe-Salpeter equation with kernel $V$.
}\label{figure1}
\end{figure}

In this section, we briefly review the Bethe-Salpeter (BS) equation \cite{BS}, and its instantaneous version, Salpeter equation \cite{salp}, both of them are the relativistic dynamic equation
describing a bound state. Figure \ref{figure1} show that a quark and an anti-quark are bounded by the kernel $V$ to a meson within the framework of the BS equation, where the index of color is ignored. According to this figure, the BS equation can be written as \cite{BS}
\begin{equation}
\chi_{_P}(q)=\frac{i}{\slashed{p}_{1}-m_{1}}i\int\frac{d^{4}k}{(2\pi)^{4}}
V(P,k,q)\chi_{_P}(k)\frac{i}{-\slashed{p}_{2}-m_{2}}\;, \label{eq2-1}
\end{equation}
where $\chi_{_P}(q)$ is the relativistic wave function of the bound state, $V(P,k,q)$ is the interaction kernel, $m_1,m_2$ are the masses of the constitute quark and anti-quark. The relation between the whole momentum $P$, the relative momentum $q$, and quark momenta $p_1$ and $p_2$ are $p_{1}={\alpha}_{1}P+q,~ p_{2}={\alpha}_{2}P-q,$
where ${\alpha}_{1}=\frac{m_{1}}{m_{1}+m_{2}}$ and ${\alpha}_{2}=\frac{m_{2}}{m_{1}+m_{2}}$.

As a four-dimensional integral equation, the BS equation is hard to be solved, so many approximate versions are proposed. Among them, the famous Salpeter equation \cite{salp} is the instantaneous approach of BS equation, which is suitable for a bound state with one heavy quark, very good for a doubly heavy quarkonium.

When we take the instantaneous approximation, in the center of mass frame of the bound state, the kernel $V(P,k,q)$ becomes to $ V(k_{\perp},q_{\perp})$, where $q^{\mu}_{\perp}\equiv q^{\mu}-q^{\mu}_{\parallel}$, and
$q^{\mu}_{\parallel}\equiv (P\cdot q/M^{2})P^{\mu}$,
$M$ is the mass of meson. We introduce the three-dimensional wave function
\begin{equation}
\Psi_{P}(q^{\mu}_{\perp})\equiv i\int
\frac{dq_{_P}}{2\pi}\chi(q^{\mu}_{\parallel},q^{\mu}_{\perp}),
\end{equation}
where $q_{_P}=\frac{(P\cdot q)}{M}$.
Then the BS Eq. (\ref{eq2-1}) can be rewritten as
\begin{equation}
\chi(q_{\parallel},q_{\perp})=S_{1}(p_{1})\eta(q_{\perp})S_{2}(p_{2}),
\label{eq2-2}
\end{equation}
where we have defined
\begin{equation}
\eta(q^{\mu}_{\perp})\equiv\int\frac{dk^3_{\perp} }{(2\pi)^{3}}
V(k_{\perp},q_{\perp},s)\Psi_{P}(k^{\mu}_{\perp}),
\end{equation}
$S_{1}(p_{1})$ and $S_{2}(p_{2})$ are the propagators
\begin{equation}
S_{i}(p_{i})=\frac{\Lambda^{+}_{i}(q_{\perp})}{(-1)^{i+1} q_{_P} +\alpha_{i}M-\omega_{i}+i\varepsilon}+
\frac{\Lambda^{-}_{i}(q_{\perp})}{(-1)^{i+1} q_{_P}+\alpha_{i}M+\omega_{i}-i\varepsilon},
\label{eq2-3}
\end{equation}
where $\omega_{i}=\sqrt{m_{i}^{2}+q^{2}_{_T}}$ with $q_{_T}=\sqrt{q^{2}_{_P}-q^{2}}=\sqrt{-q^{2}_{\perp}}$. The project operator is
$\Lambda^{\pm}_{i}(q_{\perp})= \frac{1}{2\omega_{i}}\left[ \frac{\slashed{P}}{M}\omega_{i}\pm
(-1)^{i+1} (m_{i}+\slashed{q}_{\perp})\right]$, where $i=1$ and $2$ for quark and anti-quark, respectively.

With the definition $\Psi^{\pm\pm}_{P}(q_{\perp})\equiv \Lambda^{\pm}_{1}(q_{\perp})
\frac{\slashed{P}}{M}\Psi_{P}(q_{\perp}) \frac{\slashed{P}}{M} \Lambda^{{\pm}}_{2}(q_{\perp})$, the wave function can be rewritten as four terms
\begin{equation}
\Psi_{P}(q_{\perp})=\Psi^{++}_{P}(q_{\perp})+
\Psi^{+-}_{P}(q_{\perp})+\Psi^{-+}_{P}(q_{\perp})
+\Psi^{--}_{P}(q_{\perp}), \label{eq2-4}
\end{equation}
where $\Psi^{++}_{P}(q_{\perp})$ and $\Psi^{--}_{P}(q_{\perp})$ are usually called as the positive and negative energy wave functions.

Integrating over $q_{_P}$ on both sides of Eq. (\ref{eq2-2}), the Salpeter equation \cite{salp} is obtained:
\begin{equation}
\Psi_{P}(q_{\perp})=\frac{
\Lambda^{+}_{1}(q_{\perp})\eta(q_{\perp})\Lambda^{+}_{2}(q_{\perp})}
{(M-\omega_{1}-\omega_{2})}- \frac{
\Lambda^{-}_{1}(q_{\perp})\eta(q_{\perp})\Lambda^{-}_{2}(q_{\perp})}
{(M+\omega_{1}+\omega_{2})}.\label{salpe}
\end{equation}
The upper Salpeter equation can be also described as four equations by using the project operators:
\begin{equation}
(M-\omega_{1}-\omega_{2})\Psi^{++}_{P}(q_{\perp})=
\Lambda^{+}_{1}(q_{\perp})\eta(q_{\perp})\Lambda^{+}_{2}(q_{\perp})\;,\label{positive}
\end{equation}
\begin{equation}
(M+\omega_{1}+\omega_{2})\Psi^{--}_{P}(q_{\perp})=-
\Lambda^{-}_{1}(q_{\perp})\eta(q_{\perp})\Lambda^{-}_{2}(q_{\perp})\;,\label{negative}
\end{equation}
\begin{equation}
\Psi^{+-}_{P}(q_{\perp})=0,~\Psi^{-+}_{P}(q_{\perp})=0\;.
\label{pone}
\end{equation}

The normalization condition is
\begin{equation}
\int\frac{q_{_T}^2dq_{_T}}{2{\pi}^2}Tr\left[\overline\Psi^{++}_{P}
\frac{{/}\!\!\!
{P}}{M}\Psi^{++}_{P}\frac{{/}\!\!\!{P}}{M}-\overline\Psi^{--}_{P}
\frac{{/}\!\!\! {P}}{M}\Psi^{--}_{P}\frac{{/}\!\!\!
{P}}{M}\right]=2M\;. \label{nor}
\end{equation}

\section{Kernels in different gauges}

\begin{figure}[htb]
\begin{picture}(470,600)(125,490)
\put(0,0){\includegraphics{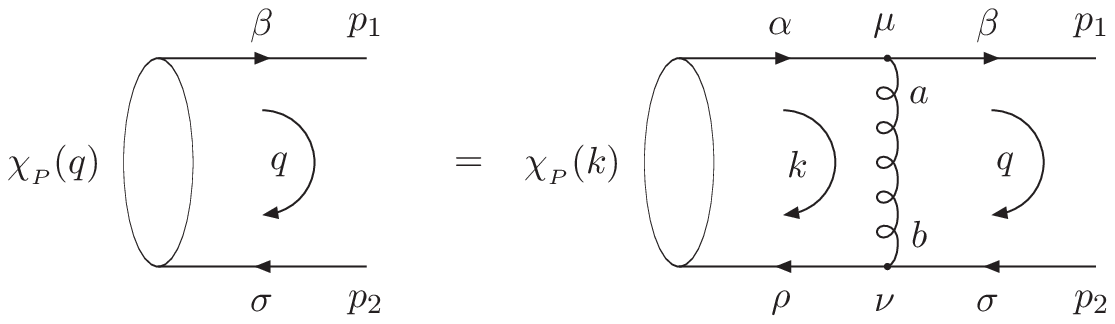}}
\end{picture}
\caption{Bethe-Salpeter equation with one gluon exchange kernel.
}\label{figure2}
\end{figure}

In our calculation, the interaction kernel can be written as $V=V_s+V_v$, where $V_s$ is the scalar potential, and $V_v$ the vector potential. In this paper, the long range linear confining scalar potential $\lambda r$ for $V_s$ is chosen, which can not be derived from the perturbative QCD, but suggested by the Color confinement of quarks in bound state and by Lattice QCD calculations \cite{godfrey}.
In a potential model, a free constant $V_0$ is usually introduced in the scalar interaction to fit data, $V_s=\lambda r+V_0$.

The vector part $V_v$ can be derived using the perturbative QCD within the BS equation.
In figure 2, the diagram of the BS equation with one gluon exchange is plotted. According to this Feymann diagram, the BS equation can be written as
\begin{equation}
\frac{\delta_{\beta\delta}}{\sqrt{3}}\chi_P(q)=S_1(p_1)\left(-ig\gamma^{\mu}(T^a)_{\alpha\beta}\right)
\int\frac{d^4k}{(2\pi)^4}\left[\chi_P(k)\frac{\delta_{\alpha\rho}}{\sqrt{3}}
\left(-ig\gamma^{\nu}(T^b)_{\sigma\rho}\right)S_2(-p_2)
({-i}D_{\mu\nu})\right]\delta^{ab},
\label{figeq}
\end{equation}
where $\alpha$, $\beta$, $\rho$, $\sigma$, $a$ and $b$ are the color indices, $T^a$ is the SU(3) group generator, and all the repeated indices are summed over. $-iD_{\mu\nu}\delta^{ab}$ is the gluon propagator, in Feymann gauge ($\eta=1$) or in Landau gauge ($\eta=0$), it can be written as
\begin{equation}D_{\mu\nu}=\frac{1}{(q-k)^2}
\left[g_{\mu\nu}-(1-\eta)\frac{(q-k)_{\mu}(q-k)_{\nu}}{(q-k)^2}\right],
\end{equation}
in Coulomb gauge,
\begin{equation}
D_{\mu\nu}=\frac{1}{(q-k)^2}
\left[\delta_{\mu\nu}-\frac{(q-k)_{\mu}(q-k)_{\nu}}{(q-k)^2}\right],
\end{equation}
in time-component Coulomb gauge (also called time-component Lorentz gauge \cite{lucha}),
\begin{equation}D_{\mu\nu}=\frac{g_{00}}{(q-k)^2}.
\end{equation}

After we summed over the color index,  $(T^aT^a)_{\sigma\beta}=\frac{N^2-1}{2N}\delta_{\sigma\beta}=\frac{4}{3}\delta_{\sigma\beta}$, the BS equation becomes to
\begin{equation}
(\slashed{p}_{1}-m_{1})\chi_{_P}(q)(\slashed{p}_{2}+m_{2})= \frac{i16\pi\alpha_s}{3}\int\frac{d^{4}k}{(2\pi)^{4}}\gamma^{\mu}\chi_{_P}(k)\gamma^{\nu}D_{\mu\nu}\equiv\
i\int\frac{d^{4}k}{(2\pi)^{4}}V_v\chi_{_P}(k),
\end{equation}
where $\alpha_s=\frac{g^2}{4\pi}$ is the running coupling constant, in momentum space and at one-loop level, it can be written as
$$\alpha_s(\vec{q})=\frac{12\pi}{33-2N_f}\frac{1}
{\log (e+\frac{{\vec{q}}^2}{\Lambda^{2}_{QCD}})},$$
and $e=2.71828$, $N_f$ is the active quark number, $N_f=3$ for charmonium and heavy-light system, $N_f=4$ for bottomonium.

In order to avoid the infrared divergence in momentum space and incorporate the screening effects, we add an exponential factor $e^{-\alpha r}$ to the potential \cite{laermann}, $\lambda r \rightarrow \frac{\lambda}{\alpha}(1-e^{-\alpha r})$.
In the instantaneous approach, the four dimensional quantity $q-k$ becomes to $q_{\perp}-k_{\perp}$. So in the momentum space and the rest frame of the bound state, the scalar potential takes the form:
$$V_s(q_{\perp}-k_{\perp})=-\left(\frac{\lambda}{\alpha}+V_0\right)
\delta^3(q_{\perp}-k_{\perp})+\frac{\lambda}{\pi^2}
\frac{1}{\left((q_{\perp}-k_{\perp})^2-{\alpha}^2\right)^2}.$$

\section{Relativistic wave function}

Although Salpeter equation is the relativistic dynamic equation, it cannot by itself provide the information about the wave function. To keep the advantage of relativistic equation, we need to provide a form of the relativistic wave function as input. A familiar form of the non-relativistic wave function for the pseudoscalar $0^-$ meson is $\Psi(q_{\perp})=(\slashed{P}+M)\gamma_5\varphi(q_{\perp})$, while
a relativistic wave function form is \cite{wang}
\begin{equation}
\Psi(q_{\perp})=\left[\slashed{P}\varphi_1(q_{\perp})+M\varphi_2(q_{\perp})
+\slashed{q}_{\perp}\varphi_3(q_{\perp})
+\frac{\slashed{P}\slashed{q}_{\perp}}{M}\varphi_4(q_{\perp})\right]\gamma_5,
\end{equation}
where the radial wave function $\varphi_i$ is a function of $q^2_{\perp}$, so no $q^2_{\perp}$ and higher order terms appear in the wave function. Because of the instantaneous approach $P\cdot q_{\perp}=0$, also no $P\cdot q_{\perp}=0$ terms in the wave function. We note that the $q$ dependent terms are relativistic, if we delete them, and set $\varphi_1(q_{\perp})=\varphi_2(q_{\perp})$, then the wave function reduces to the non-relativistic case.

Not all the four radial wave functions $\varphi_i$ are independent, equation Eq.(10) provide the connections
$$\varphi_3(q_{\perp})=\frac{M(\omega_2-\omega_1)}{(m_1\omega_2+m_2\omega_1)},
~~\varphi_4(q_{\perp})=-\frac{M(\omega_1+\omega_2)}{(m_1\omega_2+m_2\omega_1)}.$$
We have two equations Eqs.(8-9), and two unknown radial wave functions $\varphi_1(q_{\perp})$ and $\varphi_2(q_{\perp})$, after taking trace, we can solve the coupled equations as the eigenvalue problem, and obtain the mass spectrum and  radial wave functions numerically.

\section{Results and discussion}

In the calculation, we set the quark masses $m_b=4.96~\mathrm{GeV}$, $m_c=1.62~\mathrm{GeV}$, $m_s=0.50~\mathrm{GeV}$, $m_u=0.305~\mathrm{GeV}$, and  parameters $\alpha=0.06$ GeV, $\Lambda_{\rm QCD}=0.16$ GeV, $\lambda=0.18$ GeV$^2$. To compare the results between different gauges, we choose the same parameters for all the gauges used here, only vary $V_0$ to fit the ground state mass.  With these parameters, we solve the full Salpeter equations for heavy pseudoscalars with kernels in different gauges, and present numerical mass spectra.

In table I, we show the mass spectra of $D^0(nS)$ using four different kernels, and the experimental data from Particle Data Group \cite{pdg} are also shown. As shown in Sec.III, the kernels are obtained by using linear scalar potential plus one gluon exchange vector potential, where the gluon propagator has been set in Feynman, Landau, Coulomb, and time-component Coulomb gauge, separately. In the following tables, the columns of ¡®Feynman¡¯, ¡®Landau¡¯, ¡®Coulomb¡¯, and ¡®time-component Coulomb¡¯ are referred to different kernels in corresponding gauges.

\begin{table} \caption{Mass spectra of $D^0(nS)$ in units of MeV with different gauges.}
\begin{center}
\begin{tabular}{|c|c|c|c|c|c|} \hline\hline
~~~~&~~Feynman~~&~~Landau~~&~~Coulomb~~&~time-component Coulomb~&~Exp~\cite{pdg}\\\hline\hline
~~$1S$~~&~~1864.8~~&~~1864.8~~&~~1864.7~~&~~1864.5~~&~~1864.83~~\\\hline
~~$2S$~~&~~2611.2~~&~~2420.9~~&~~2573.9~~&~~2309.2~~&~~2539.4~\cite{2550}~\\\hline
~~$3S$~~&~~2898.6~~&~~2697.4~~&~~2750.5~~&~~2564.3~~&~~\\\hline
~~$4S$~~&~~3071.4~~&~~2877.5~~&~~3016.8~~&~~2737.6~~&~~\\\hline\hline
\end{tabular}\label{tab1}
\end{center}
\end{table}

The $D(2S)$ has been discovered by two experiments, the BaBar Collaboration found its mass is $2539.4\pm4.5\pm6.8$ MeV, LHCb Collaboration \cite{2580} obtained $2579.5\pm3.4\pm5.5$ MeV. Our result in Coulomb gauge consist with experimental data, while Feynman gauge gives a heavier mass, Landau and time-component Coulomb gauges provide a smaller one.

\begin{table} \caption{Mass spectra of $D_s(nS)$ in units of MeV using different gauge kernels.}
\begin{center}
\begin{tabular}{|c|c|c|c|c|c|} \hline\hline
~~~&~~Feynman~~&~~Landau~~&~~Coulomb~~&~time-component Coulomb~&~~Exp~\cite{pdg}~\\\hline\hline
~~$1S$~~&~~1968.3~~&~~1968.3~~&~~1968.2~~&~1968.0~&~~1968.34~~\\\hline
~~$2S$~~&~~2811.6~~&~~2582.6~~&~~2754.5~~&~2442.4~&~~~~\\\hline
~~$3S$~~&~~3141.6~~&~~2894.4~~&~~3074.9~~&~2727.1~&~~\\\hline
~~$4S$~~&~~3350.2~~&~~3105.3~~&~~3275.5~~&~2928.1~&~~\\\hline\hline
\end{tabular}\label{tab2}
\end{center}
\end{table}

\begin{table} \caption{Mass spectra of $B^{\pm}(nS)$ in units of MeV using different gauge kernels.}
\begin{center}
\begin{tabular}{|c|c|c|c|c|c|} \hline\hline
~~~&~~Feynman~~&~~Landau~~&~~Coulomb~~&~time-component Coulomb~&~~Exp~~\cite{pdg}~\\\hline\hline
~~$1S$~~&~~5279.3~~&~~5279.3~~&~~5279.5~~&~5279.6~&~5279.32~\\\hline
~~$2S$~~&~~5799.5~~&~~5727.1~~&~~5793.2~~&~5692.9~&~~\\\hline
~~$3S$~~&~~6067.0~~&~~5978.8~~&~~6052.3~~&~5932.8~&~~\\\hline
~~$4S$~~&~~6247.7~~&~~6153.4~~&~~6225.1~~&~6100.6~&~~\\\hline\hline
\end{tabular}\label{tab3}
\end{center}
\end{table}

The mass spectra of $D_s$, $B$ and $B_s$ are given in Tables II-IV. At present, there are no candidates of these radial excited states in experiment. From these three tables, similar to the $D$ meson, the kernel in time-component Coulomb gauge always gives the smallest mass splitting, Feynman and Coulomb gauges provide similar largest mass splitting, and Landau gauge result is in the middle.

\begin{table} \caption{Mass spectra of $B_s(nS)$ in units of MeV using different gauge kernels.}
\begin{center}
\begin{tabular}{|c|c|c|c|c|c|} \hline\hline
~~~&~~Feynman~~&~~Landau~~&~~Coulomb~~&~time-component Coulomb~&~~~Exp~~\cite{pdg}~\\\hline\hline
~~$1S$~~&~~5366.4~~&~~5366.3~~&~~5363.3~~&~5366.9~&~~5366.89~~\\\hline
~~$2S$~~&~~5945.4~~&~~5855.2~~&~~5943.2~~&~5810.8~&~~\\\hline
~~$3S$~~&~~6246.8~~&~~6137.6~~&~~6239.1~~&~6079.3~&~~\\\hline
~~$4S$~~&~~6454.9~~&~~6338.4~~&~~6440.3~~&~6272.6~&~~\\\hline\hline
\end{tabular}\label{tab4}
\end{center}
\end{table}

\begin{table} \caption{Mass spectra of $B_c(nS)$ in units of MeV using different gauge kernels.}
\begin{center}
\begin{tabular}{|c|c|c|c|c|c|} \hline\hline
~~&~~Feynman~~&~~Landau~~&~~Coulomb~~&~time-component Coulomb~&~~Exp~~\cite{pdg}~\\\hline\hline
~~$1S$~~&~~6274.5~~&~~6274.1~~&~~6274.1~~&~6274.7~&~~6274.9~~\\\hline
~~$2S$~~&~~6972.1~~&~~6833.3~~&~~6983.8~~&~6746.8~&~~6842~\cite{bc2s}~\\\hline
~~$3S$~~&~~7320.8~~&~~7148.5~~&~~7336.9~~&~7043.3~&~~\\\hline
~~$4S$~~&~~7568.0~~&~~7386.8~~&~~7581.7~~&~7268.0~&~~\\\hline\hline
\end{tabular}\label{tab5}
\end{center}
\end{table}

The meson $B_c(2S)$ is first observed by ATLAS Collaboration \cite{bc2s}, and the mass is measured to be $6842\pm4\pm5$ MeV. Later, CMS Collaboration reported the observation of the $B_c(2S)$ and $B^*_c(2S)$, and the mass peak of these two mesons is located at $6871.2\pm1.2\pm0.8\pm0.8$ MeV, the mass difference between them is $29.0\pm1.5\pm0.7$ MeV. Recently, LHCb Collaboration confirmed the observation of the $B_c(2S)$ and $B^*_c(2S)$, they reported the mass of $B^*_c(2S)$ is $6841.2\pm0.6\pm0.1\pm0.8$ MeV, while $B_c(2S)$ is $6872.1\pm1.3\pm0.1\pm0.8$ MeV. Theoretically, our Landau gauge result is consistent with data, Feynman and Coulomb gauges provide $100$ MeV larger, while, time-component Coulomb result is $100$ MeV smaller.

\begin{table} \caption{Mass spectra of $\eta_c(nS)$ in units of MeV using different gauge kernels.}
\begin{center}
\begin{tabular}{|c|c|c|c|c|c|} \hline\hline
~~&~~Feynman~~&~~Landau~~&~~Coulomb~~&~time-component Coulomb~&~~Exp~~\cite{pdg}~\\\hline\hline
~~$1S$~~&~~2983.3~~&~~2983.9~~&~~2983.7~~&~2984.0~&~~2983.9~~\\\hline
~~$2S$~~&~~3874.2~~&~~3628.9~~&~~3843.3~~&~3472.0~&~~3637.6~~\\\hline
~~$3S$~~&~~4253.5~~&~~3972.8~~&~~4219.3~~&~3780.4~&~~\\\hline
~~$4S$~~&~~4508.3~~&~~4217.8~~&~~4468.6~~&~4010.2~&~~\\\hline\hline
\end{tabular}\label{tab6}
\end{center}
\end{table}

\begin{table} \caption{Mass spectra of $\eta_b(nS)$ in units of MeV using different gauge kernels.}
\begin{center}
\begin{tabular}{|c|c|c|c|c|c|} \hline\hline
~~&~~Feynman~~&~~Landau~~&~~Coulomb~~&~time-component Coulomb~&~~Exp~~\cite{pdg}~\\\hline\hline
~~$1S$~~&~~9399.5~~&~~9399.7~~&~~9399.9~~&~9400.8~&~~9399.0~~\\\hline
~~$2S$~~&~~10152.7~~&~~10004.7~~&~~10215.2~~&~9898.8~&~~9999.0~\cite{mizuk}~~\\\hline
~~$3S$~~&~~10516.1~~&~~10333.4~~&~~10590.2~~&~10190.2~&~~\\\hline
~~$4S$~~&~~10768.7~~&~~10569.4~~&~~10842.7~~&~10417.6~&~~\\\hline\hline
\end{tabular}\label{tab7}
\end{center}
\end{table}

Mass spectra of charmonium and bottomonium are shown in Table VI and Table VII. For charmonium, the first radial excited state $\eta_c(2S)$ has been detected, whose mass is $3637.6\pm1.2$ MeV in PDG. Our prediction of Landau gauge is consistent with data, while the theoretical results of Feynman and Coulomb gauges are about $200$ MeV larger, and time-component Coulomb gauge is about $160$ MeV smaller. For bottomonium, $\eta_b(2S)$ is also discovered by Belle Collaboration, its mass is $9999.0\pm3.5^{+2.8}_{-1.9}$ MeV \cite{mizuk}, which is consistent with our Landau gauge result. As usual, the predictions of Feynman and Coulomb gauges are $150-210$ MeV larger, and time-component Coulomb gauge is $100$ MeV smaller.

In summary, we solved the instantaneous BS equation, which is also called Salpeter equation for heavy pseudoscalars, using four different kernels caused by different gauges. We found in the current choice of parameters, the prediction of Landau gauge is more consistent with experimental data. But we should point out that, if another set of parameters is chosen, the conclusion may be different, this is also the reason why different potentials and gauges existing in literature, but the following conclusion is remain unchanged,
the mass splitting in the time-component Coulomb gauge is always the smallest one, while Landau gauge gives a middle size mass splitting, and Feynman and Coulomb gauges provide the similar largest mass splitting.

\section{Acknowledgments}

This work was supported in part by the National Natural Science Foundation of China (NSFC) under Grant No.~11575048 and No.~11535002.

\end{document}